\documentclass[prd,twocolumn,superscriptaddress,floatfix,nofootinbib]{revtex4-2}
\pdfoutput=1 
\usepackage[T1]{fontenc} 
\usepackage{amsmath,amssymb,amsfonts,amsthm}
\usepackage{graphicx}
\usepackage[usenames,dvipsnames]{color}
\usepackage{enumitem}
\usepackage{verbatim}
\usepackage[percent]{overpic}
\usepackage{rotating}
\usepackage{hyperref}
\usepackage{array}
\usepackage{mathtools}
\usepackage{lmodern}
\usepackage[normalem]{ulem}
\usepackage{braket}
\usepackage{tensor}
\usepackage{dsfont}
\usepackage{bm}
\usepackage{tikz}

\usepackage{mathrsfs}

\usetikzlibrary{decorations.pathmorphing}
\usepackage{CJKutf8}

\usetikzlibrary{positioning}
\usetikzlibrary{calc,through,backgrounds}

\theoremstyle{definition}
\newtheorem{definition}{Definition}[section]
\newtheorem{theorem}{Theorem}[section]

\newcommand{\kako}[1]{\left( #1 \right)}
\newcommand{\kagikako}[1]{\left[ #1 \right]}
\newcommand{\ts}[1]{ _{\text{#1}} }
\newcommand{\Bigkako}[1]{\Big( #1 \Big)}

\newcommand{\Bigkagikako}[1]{\Big[ #1 \Big]}

\newcommand{\erfi}{\text{erfi}}

\allowdisplaybreaks

\DeclareMathOperator{\Tr}{Tr}

\newcommand{\dd}{\text{d}}

\newcommand{\sx}{\mathsf{x}}

\newcommand{\ii}{\mathsf{i}}

\newcommand{\kk}{|\bm{k}|}

\begin{document}

\title{Entangled detectors nonperturbatively harvest mutual information}


\author{Kensuke Gallock-Yoshimura}
\email{kgallock@uwaterloo.ca}
\affiliation{Department of Physics and Astronomy, University of Waterloo, Waterloo, Ontario, N2L 3G1, Canada}


\author{Robert B. Mann}
\email{rbmann@uwaterloo.ca}
\affiliation{Department of Physics and Astronomy, University of Waterloo, Waterloo, Ontario, N2L 3G1, Canada}
\affiliation{Perimeter Institute for Theoretical Physics,  Waterloo, Ontario, N2L 2Y5, Canada}

\begin{abstract}
We investigate how entangled inertial Unruh-DeWitt detectors are affected by interaction with a quantum field using a nonperturbative method. 
Inertial detectors in a $(3+1)$-dimensional Minkowski spacetime 
with instantaneous switching ($\delta$-switching) experience degradation of their initial entanglement as their coupling strength with a scalar field increases. 
Somewhat surprisingly, initially separable or weakly entangled detectors  can extract mutual information from the vacuum. 
We also find that entanglement degradation is not reduced if communication via the field is possible; rather this only changes the manner in which entanglement is degraded.
\end{abstract}

\maketitle
\flushbottom

\section{Introduction}
The fact that quantum fields are entangled \cite{summers1985bell, summers1987bell} even in the vacuum state has attracted increasing attention over the past decade. Originally, Valentini \cite{Valentini1991nonlocalcorr} and subsequently Reznik \textit{et al.} \cite{reznik2003entanglement, reznik2005violating} showed that particle detectors can become  entangled by extracting  correlations in the field, even without directly exchanging quanta. 
This kind of entanglement extraction from the field is called \textit{entanglement harvesting}, and  has been examined extensively \cite{Steeg2009, pozas2015harvesting, salton2015acceleration, smith2016topology, kukita2017harvesting, Simidzija.Nonperturbative, Simidzija2018no-go, henderson2018harvesting, ng2018AdS, henderson2019entangling, cong2019entanglement, Tjoa2020vaidya, cong2020horizon, Xu.Grav.waves.PRD.102.065019, Ken.Freefall.PhysRevD.104.025001}. 
The general approach has been to consider initially separable Unruh-DeWitt (UDW) detectors \cite{Unruh1979evaporation, DeWitt1979} interacting with a scalar field for a finite time. As long as no angular momentum is exchanged, this is a good approximation to the light-matter interaction
\cite{PhysRevD.87.064038,PhysRevA.89.033835,Pozas-Kerstjens:2016}.

While entanglement harvesting is about extracting entanglement from the field, there is a phenomenon called \textit{entanglement degradation}, in which the amount of a prepared entanglement is   reduced due to the field. 
Typically, the Unruh and Hawking effects have been of primary interest \cite{PhysRevLett.91.180404, PhysRevLett.95.120404, PhysRevD.78.065015, PhysRevA.80.032315, WANG2010202, PhysRevD.82.064006, PhysRevA.74.032326, FermionicEMM, PhysRevA.82.032324}: maximally entangled field modes from an inertial observer's viewpoint will be less entangled from an accelerating observer's point of view because of the noisy Unruh effect. 
In addition, it is well known that entanglement of bosonic fields vanishes in  the large acceleration limit, while that of fermionic fields survives  \cite{PhysRevLett.95.120404,PhysRevA.74.032326, FermionicEMM, Montero.arbitrarySpin.PhysRevA.84.012337}. 
In the context of the UDW detectors, initially entangled, noninertial detectors lose some (or all) of the correlation after interacting with the field \cite{PhysRevA.80.032315, PhysRevD.78.125025, Doukas.orbit.PhysRevA.81.062320}. 

Our interest is whether or not an initial entanglement between two detectors is always degraded. 
While initially entangled detectors experience entanglement degradation, it is conceivable that they could also gain entanglement from the vacuum according to the entanglement harvesting protocol. 
Moreover, communication between the detectors enhances entanglement in the harvesting scenario, but it is not clear how communication plays a role in entanglement degradation. 
In fact, it has been shown \cite{PhysRevD.78.125025}   that entanglement can be revived for a while even after detectors are completely disentangled. 
It is therefore of interest   to see in what circumstances  initially entangled detectors gain or lose entanglement, or more generally gain or lose correlation.

To this end, we nonperturbatively analyze two \textit{inertial} UDW detectors coupled to a scalar field in $(3+1)$-dimensional Minkowski spacetime, while paying attention to their communication channels. 
We will make use of a nonperturbative analysis \cite{Simidzija.Nonperturbative} in which the UDW detectors are assumed to switch on and off instantaneously, modeled by a Dirac $\delta$-function.
Such a method enables us to obtain a density matrix of detectors that clearly shows how communication and the initial state contribute to entanglement degradation. 

Our main findings are the following. 
(a) Detectors experience entanglement degradation as their coupling with the field gets stronger if  only one of or both of the detectors interact with the field. 
(b) Although mutual information (classical and quantum correlations) between the detectors exhibits similar behavior to entanglement, weakly entangled detectors can actually gain mutual information from the vacuum, even without  communication. 
(c) Communication does not enhance either entanglement or mutual information. 
However distinct initial entangled states with the same amount of entanglement show different degradation properties, which purely originate from communication. 

Our  paper is organized as follows. 
In Sec. \ref{subsec:Time-evolution operator} we review the UDW detectors with the ``delta switching" (i.e., interacting instantaneously).  
Then we derive the density matrix of the detectors after the interaction in \ref{subsec:Density matrix}. 
To be more specific, it is a density matrix of two initially entangled, inertial detectors in an $(n+1)$-dimensional Minkowski spacetime. 
We then further assume $n=3$ and that the detectors' shapes are both specified by a Gaussian smearing function. 
We give measures of correlations, concurrence, and mutual information in Sec.  \ref{subsec:Correlation measure}. 
Results are shown in Sec. \ref{sec: results}. 
In particular, we consider two cases: only one detector interacts with the field (\ref{subsec:Single detector interacts with the field}) and both detectors interact (\ref{subsec:Two detectors interact with the field}) after the entangled detectors are prepared. 
Finally we conclude in Sec. \ref{sec: conclusion}. 
Throughout the paper, we use natural units $\hbar =c=1$, and denote spacetime points by $\sx=(t,\bm{x})$.

\section{UDW detectors with delta switching}
\label{sec: UDW detectors with delta switching}

\subsection{Time-evolution operator}
\label{subsec:Time-evolution operator}
Consider two UDW detectors A and B interacting with a local quantum field $\hat \phi(\sx )$. 
The interaction Hamiltonian of a linearly coupled detector-$j\in \{ \text{A,B} \}$ is given by
\begin{align}
    \hat H_j^{\tau_j}(\tau_j) 
    &= \lambda_j \chi_j(\tau_j) \hat \mu_j(\tau_j) 
    \otimes 
    \int \dd^nx\,F_j(\bm{x}-\bm{x}_j)
    \hat \phi(\sx_j(\tau_j)), 
\end{align}
where $\tau_j, \lambda_j, \chi_j(\tau_j)$, and $F_{j}(\bm{x}-\bm{x}_j)$ are the proper time, the coupling constant, the switching function, and the smearing function of detector-$j$, respectively. 
The superscript $\tau_j$ indicates that it is a generator of time-translations with respect to $\tau_j$. 
The switching function $\chi_j(\tau_j)$ determines the time-dependence of coupling, and the smearing function $F_j(\bm{x}-\bm{x}_j)$ represents the size and the shape of a detector whose center of mass is located at $\bm{x}_j$. 
The operator $\hat \mu_j (\tau_j)$ is the monopole moment given by
\begin{align}
    \hat\mu_j(\tau_j) &=\ket{e_j}\!\bra{g_j}e^{\ii\Omega_j\tau_j } + \ket{g_j}\!\bra{e_j}e^{-\ii\Omega_j\tau_j }\,,
\end{align}
where $\ket{g_j}, \ket{e_j}$ are the respective ground and excited states of detector-$j$, and $\Omega_j$ is the energy gap between these two states. 

Let us employ the  switching function defined by \cite{Simidzija.Nonperturbative}
\begin{align}
    \chi_j(\tau_j)= 
    \eta_j \delta(\tau_j - \tau_{j,0}),  
\end{align}
which we refer to as delta switching, 
where $\tau_{j,0}$ is the proper time when detector-$j$ is switched, and $\eta_j$ is a constant, which has units of time, so that 
\begin{align}
    \int_{-\infty}^\infty \chi_j(\tau_j)\dd \tau_j=\eta_j
\end{align}
is constant. 

By making use of delta switching, we can analyze the properties of detectors nonperturbatively. Let us introduce a coordinate system $(t,\bm{x})$ that specifies the positions of both detectors. 
In our analysis, such a coordinate system will be Cartesian coordinates in Minkowski spacetime. 
Since the time component $t$ can specify time for both detectors, the proper times $\tau_j$ can be written in terms of $t$, i.e., $\tau_j(t)$. 
By using this common time $t$, the time-evolution operator $\hat U\ts{I}$ in the interaction picture can be written as follows. 
\begin{align}
    \hat U\ts{I}&=
    \mathcal{T}_t 
    \exp
    \kagikako{
        -\ii \int_{\mathbb{R}} \dd t\,\hat H^t\ts{I} (t)
    }  ,
\end{align}
where $\mathcal{T}_t$ is time-ordering symbol with respect to the common time $t$, and the Hamiltonian $\hat H^t\ts{I} (t)$ is \cite{EMM.Relativistic.quantum.optics, Tales2020GRQO}
\begin{align}
    \hat H^t\ts{I} (t)&=
    \dfrac{\dd \tau\ts{A}}{\dd t}
            \hat H\ts{A}^{\tau\ts{A}}(\tau\ts{A}(t)) 
            +
            \dfrac{\dd \tau\ts{B}}{\dd t}
            \hat H\ts{B}^{\tau\ts{B}}(\tau\ts{B}(t)). 
\end{align}
Assuming that detector A switches earlier than B, namely, $t(\tau\ts{A,0})\leq t(\tau\ts{B,0})$, the delta switching enables us to write the time-evolution operator $\hat U\ts{I}$ as \cite{Simidzija.Nonperturbative}
\begin{align}
    \hat U\ts{I}
    &= 
    \exp 
    \kagikako{
        \hat \mu\ts{B}(\tau\ts{B,0})\otimes \hat Y\ts{B}(\tau\ts{B,0})
    }
    \exp 
    \kagikako{
        \hat \mu\ts{A}(\tau\ts{A,0})\otimes \hat Y\ts{A}(\tau\ts{A,0})
    },\label{eq: time-evolution operator}
\end{align}
where 
\begin{align}
    \hat Y_j (\tau_{j,0})\coloneqq 
    -\ii \lambda_j \eta_j 
    \dfrac{\dd \tau_j}{\dd t} \bigg|_{\tau_{j,0}}
    \int \dd^nx\,F_j(\bm{x}-\bm{x}_j)
    \hat \phi(\sx_j(\tau_{j,0})). \label{eq:Y}
\end{align}
The operator $\hat Y_j$ is essentially a smeared field operator at the switching-time $t(\tau_{j,0})$. 
We finally define a commutator $\Theta$ and an anti-commutator $\omega$ as 
\begin{align}
    \Theta&\coloneqq  -\ii \bra{0} [ \hat Y\ts{A}, \hat Y\ts{B} ] \ket{0}, \\
    \omega&\coloneqq 2 \bra{0} \{  \hat Y\ts{A}, \hat Y\ts{B} \} \ket{0} .
\end{align}
These quantities $\Theta$ and $\omega$ can be considered as the Pauli-Jordan and Hadamard distributions \cite{birrell1984quantum}, respectively. 
The quantity $\Theta$ is nonzero whenever detectors can send and receive signals but vanishes otherwise (e.g., when the support of smeared detectors are spatially separated), and so it provides a measure of   communication between detectors. 
On the other hand, $\omega$ can be nonzero even when detectors cannot communicate with each other. 
For this reason, it is associated with  correlations between noncommunicating detectors. 
For example, only the Hadamard distribution contributes to entanglement harvesting with spacelike separated detectors.

\subsection{Density matrix}
\label{subsec:Density matrix}
Suppose the detectors are initially in the following state
\begin{align} 
    \ket{\psi}= \alpha \ket{g\ts{A}g\ts{B}} + \sqrt{ 1-\alpha^2 } e^{ \ii \theta } \ket{e\ts{A} e\ts{B}}, \label{eq:state 0011}
\end{align}
where $\alpha \in [0,1]$ and $\theta \in [0, 2\pi)$ is a relative phase. 
Note that the state is separable if $\alpha =0$ or 1; otherwise it is entangled. 
In particular, two detectors are maximally entangled if $\alpha=1/\sqrt{2}$. 

Before proceeding, we remark that we could also consider 
\begin{equation}\label{altpsi}
  \ket{\tilde{\psi}} = \alpha \ket{ g\ts{A} e\ts{B} } + \sqrt{1-\alpha^2} e^{\ii \theta} \ket{ e\ts{A} g\ts{B} }
\end{equation}
to be the initial state. 
The effects of using $\ket{\tilde{\psi}}$ will be very similar to those we obtain from  $ \ket{\psi}$, since the latter state yields the former upon setting $\Omega_B \to - \Omega_B$. 
In what follows we shall comment on $\ket{\tilde{\psi}}$ as appropriate.

Assuming the quantum field is in the vacuum state $\ket{0}$, the reduced density matrix $\rho\ts{AB}$ of the detectors after  interacting with the field is   
\begin{align}
    \rho\ts{AB}
    &=\Tr_\phi [\hat U\ts{I} (\rho\ts{AB,0} \otimes \ket{0}\bra{0} ) \hat U\ts{I}^\dag] \\
    &= \left[
    \begin{array}{cccc}
    r_{11} &0 &0 &r_{14}  \\
    0 &r_{22} &r_{23} &0  \\
    0 &r_{23}^* &r_{33} &0  \\
    r_{14}^* &0 &0 & r_{44}
    \end{array}
    \right], \label{eq:final density}
\end{align}
where $\rho\ts{AB,0}=\ket{\psi}\bra{\psi}$, and we choose a   basis  $\{\, \ket{g\ts{A} g\ts{B}}, \ket{g\ts{A} e\ts{B}}, \ket{e\ts{A} g\ts{B}}, \ket{e\ts{A} e\ts{B}}\, \}$. 
We make an additional assumption that the  detectors are inertial, without  relative velocity (namely, $\frac{\dd \tau_j}{\dd t}=1$), in an $(n+1)$-dimensional Minkowski spacetime, interacting with a massless and minimally coupled scalar field. 
Then one can derive an explicit form for the elements in the density matrix $\rho\ts{AB}$ as (see Appendix \ref{appendix: density operator})
\begin{widetext}
\begin{subequations}
\begin{align}
    &r_{11}=
    \dfrac{1}{4} 
    \Big[ 1 + f\ts{A}f\ts{B} \cosh \omega + (2\alpha^2 -1)
    \big( f\ts{A} + f\ts{B}\cos (2\Theta) \big) \Big] 
    +\dfrac{1}{2} \alpha \sqrt{1-\alpha^2} f\ts{B}
    \big[ 
        f\ts{A} \sinh \omega \cos \vartheta - \sin (2\Theta) \sin \vartheta
    \big], \label{eq:r11 full} \\
    &r_{22}
    =\dfrac14 
    \kagikako{
        1-f\ts{A} f\ts{B} \cosh \omega 
        +(2\alpha^2 -1) \big( f\ts{A} - f\ts{B}\cos (2\Theta)\big)
    } 
    -\dfrac{1}{2} \alpha \sqrt{1-\alpha^2} f\ts{B}
    \big[
        f\ts{A} \sinh \omega 
        \cos \vartheta
        -\sin (2\Theta) 
        \sin \vartheta
    \big] , \\
    &r_{33}=
    \dfrac14 
    \kagikako{
        1-f\ts{A} f\ts{B} \cosh \omega 
        -(2\alpha^2 -1) \big( f\ts{A} - f\ts{B}\cos (2\Theta)\big)
    } 
    -\dfrac{1}{2} \alpha \sqrt{1-\alpha^2} f\ts{B}
    \big[
        f\ts{A} \sinh \omega 
        \cos \vartheta
        +\sin (2\Theta) 
        \sin \vartheta
    \big], \\
    &r_{44}=
    \dfrac{1}{4} 
    \Big[ 1 + f\ts{A}f\ts{B} \cosh \omega - (2\alpha^2 -1)
    \big( f\ts{A} + f\ts{B}\cos (2\Theta) \big) \Big] 
    +\dfrac{1}{2} \alpha \sqrt{1-\alpha^2} f\ts{B}
    \big[ 
        f\ts{A} \sinh \omega \cos \vartheta + \sin (2\Theta) \sin \vartheta
    \big], \\
    &r_{14}e^{\ii (\Omega\ts{A} \tau\ts{A,0} + \Omega\ts{B} \tau\ts{B,0})} \notag \\
    &=
    \dfrac{f\ts{B}}{4} 
    \Bigkagikako{
        f\ts{A} \sinh \omega 
        +\ii (2\alpha^2 -1) \sin (2\Theta)
    } 
    +\dfrac12 \alpha \sqrt{1-\alpha^2}
    \Bigkagikako{
        (1+f\ts{A}f\ts{B} \cosh \omega ) \cos \vartheta
        +
        \ii \big(f\ts{A} + f\ts{B} \cos (2\Theta) \big) \sin \vartheta
    }, \\
    &r_{23} e^{\ii (\Omega\ts{A} \tau\ts{A,0} + \Omega\ts{B} \tau\ts{B,0})} \notag \\
    &=-\dfrac{f\ts{B}}{4}
    \Big[ 
        f\ts{A} \sinh \omega + \ii (2\alpha^2 -1) \sin (2\Theta)
    \Big] 
    +\dfrac{1}{2} \alpha \sqrt{1-\alpha^2}
    \Big[ 
        (1-f\ts{A} f\ts{B} \cosh \omega) \cos \vartheta 
        +\ii \big( f\ts{A} - f\ts{B} \cos (2\Theta)\big) \sin \vartheta
    \Big]. \label{eq:r23 full}
\end{align}

\end{subequations}
\end{widetext}
Here 
\begin{align}
    &\vartheta \coloneqq
    \Omega\ts{A} \tau\ts{A,0} + \Omega\ts{B} \tau\ts{B,0} -\theta, \\
    &\beta_j(\bm{k})\coloneqq
    -\ii \dfrac{2\lambda_j \eta_j}{ \sqrt{2 |\bm{k}|} }
    \tilde{F}^*_j(\bm{k}) e^{ \ii \kk \tau_{j,0} - \ii \bm{k}\cdot \bm{x}_j  }, \label{eq:beta}\\
    &f_j\coloneqq
    \exp 
    \kako{
        -\dfrac12 \int \dd^n k\,|\beta_j(\bm{k})|^2
    }\in \mathbb{R}, \label{eq:f_j}\\
    &\Theta =
    \dfrac{\ii}{4} 
    \int \dd^n k
    \Bigkagikako{
        \beta\ts{A}^*(\bm{k}) \beta\ts{B}(\bm{k})
        -
        \beta\ts{A}(\bm{k}) \beta\ts{B}^*(\bm{k})
    } \in \mathbb{R}, \label{eq:Theta} \\
    &\omega =
    -\dfrac12 \int \dd^n k
    \Bigkagikako{
        \beta^*\ts{A}(\bm{k}) \beta\ts{B}(\bm{k})
        +
        \beta\ts{A}(\bm{k}) \beta\ts{B}^*(\bm{k})
    }\in \mathbb{R}, \label{eq:omega}
\end{align}
where $\tilde{F}_j (\bm{k})$ is the Fourier transformed smearing function. 

It is worth noting that this form lets us easily examine the behavior of correlations between the detectors. 
For instance, there are several terms that vanish for particular cases such as $\alpha=1/\sqrt{2}$, $\alpha=0,1$, or $\Theta=0$. 
In addition, we can consider a  scenario where the detectors are initially prepared in the state $\ket{\psi}$ but only one of them interacts with the field. For example, if we want only Bob to interact then we set $\lambda\ts{A}=0$, leading to $f\ts{A}=1, \Theta=\omega=0$. 

Finally let us specify the spatial profile of the detectors. 
We choose the Gaussian smearing function
\begin{align}
    F(\bm{x})=\dfrac{1}{ (\sqrt{\pi} \sigma )^n } e^{ -\bm{x}^2/\sigma^2 }, \label{eq:Gaussian smearing}
\end{align}
with a typical Gaussian width $\sigma$. 
Restricting ourselves to $(3+1)$-dimensional Minkowski spacetime, we obtain (see Appendix \ref{appendix: density operator}) 
\begin{align}
    &f_j=
    \exp
    \kako{
        -
        \dfrac{ \lambda_j^2 \eta_j^2 }{ 2\pi^2 \sigma^2 }
    }, \label{eq:eff}\\
    &\Theta 
    =\dfrac{\lambda\ts{A} \lambda\ts{B} \eta\ts{A} \eta\ts{B} }{ 4\pi^2 L \sigma } 
    \sqrt{ \dfrac{\pi}{2} } 
    \kako{
        e^{ -(\Delta \tau + L)^2/2\sigma^2 } - e^{ -(\Delta \tau - L)^2/2\sigma^2 } 
    },\label{eq:Theta2} \\
    &\omega
    =-\dfrac{ \lambda\ts{A} \lambda\ts{B} \eta\ts{A} \eta\ts{B} }{ \sqrt{2} \pi^2 L \sigma }
    \kagikako{
        D^+ 
        \kako{
            \dfrac{ \Delta \tau + L }{ \sqrt{2}\sigma }
        }
        -
        D^+ 
        \kako{
            \dfrac{ \Delta \tau - L }{ \sqrt{2}\sigma }
        }
    }, \label{eq:omega in 4D flat}
\end{align}
where $\Delta \tau \coloneqq \tau\ts{B,0} - \tau\ts{A,0}, L\coloneqq |\bm{x}\ts{B} - \bm{x}\ts{A}|$, and $D^+(x)$ is the Dawson function defined by 
\begin{align}
    D^+(x)\coloneqq
    \dfrac{ \sqrt{\pi} }{2} e^{ -x^2 } \erfi(x), \label{eq:Dawson}
\end{align}
with the imaginary error function $\erfi (x)$. 
Note that $\Delta \tau = L$ coresponds to the case when detectors are lightlike separated.

\subsection{Correlation measure}
\label{subsec:Correlation measure}

We are particularly interested in how the amount of entanglement and mutual information change after the interaction.  We shall quantify the   entanglement between the detectors  using concurrence. 

Suppose we have a density matrix $\rho\ts{AB}$ of a joint system $\mathcal{H}\ts{A}\otimes \mathcal{H}\ts{B}$. 
The concurrence of $\rho\ts{AB}$ is defined to be \cite{Wotters1998entanglementmeasure}
\begin{align}
    &\mathcal{C}(\rho\ts{AB})\coloneqq 
    \max \{ 0, w_1-w_2-w_3-w_4 \}, \\
    &\hspace{3cm}(w_1\geq w_2 \geq w_3 \geq w_4) \notag
\end{align}
where $w_i$ are the square roots of the eigenvalues of $\rho\ts{AB} \tilde{\rho}\ts{AB}$, and 
\begin{align}
    \tilde{\rho}\ts{AB}
    := (\sigma_y \otimes \sigma_y) \rho\ts{AB}^* (\sigma_y \otimes \sigma_y).
\end{align}
Here $\sigma_y$ is the Pauli-$y$ matrix. 
For the density matrix \eqref{eq:final density} we get 
\begin{align}
    w_i\in \{ \sqrt{r_{11}r_{44} }\pm |r_{14}|,\,
    \sqrt{r_{22}r_{33} }\pm |r_{23}|\, \} ,
\end{align}
and so the concurrence is either
\begin{align}
    \mathcal{C}(\rho\ts{AB})
    &= 2 \max \{ 0,\,|r_{14}|-\sqrt{r_{22} r_{33} } \}, \label{eq:concurrence r14}\\
    &(\text{when }w_1=\sqrt{r_{11}r_{44} }+ |r_{14}|) \notag \\
    \mathcal{C}(\rho\ts{AB})
    &= 2 \max \{ 0,\,|r_{23}|-\sqrt{r_{11} r_{44} } \}. \label{eq:concurrence r23} \\
    &(\text{when }w_1=\sqrt{r_{22}r_{33} }+ |r_{23}|)\notag
\end{align}
If the initial state is \eqref{eq:state 0011} then we always get \eqref{eq:concurrence r14}.  
On the other hand, if we use \eqref{altpsi} as the initial state, then the concurrence is given by \eqref{eq:concurrence r23}.

Let us now define mutual information, which is a measure of total correlation including classical and quantum. 
The mutual information $I(\rho\ts{AB})$ between detectors A and B is defined to be \cite{nielsen2000quantum}
\begin{align}
    I(\rho\ts{AB})\coloneqq S(\rho\ts{A}) + S(\rho\ts{B}) - S(\rho\ts{AB}), 
\end{align}
where $\rho\ts{A}=\Tr\ts{B}[\rho\ts{AB}]$ (and vice-versa), and $S(\rho)\coloneqq -\Tr [\rho \ln \rho]$ is the von Neumann entropy. 

We can evaluate mutual information explicitly from the following density matrices.
\begin{align}
    \rho\ts{A}&=
    \left[
    \begin{array}{cc}
    P\ts{A}^g &0  \\
    0 & P\ts{A}^e
    \end{array}
    \right],~
    \rho\ts{B}=
    \left[
    \begin{array}{cc}
    P\ts{B}^g &0  \\
    0 & P\ts{B}^e
    \end{array}
    \right],
\end{align}
where $P_j^g$ and $P_j^e$ are the probabilities of detector-$j$ being in ground and excited states, respectively, and they read
\begin{align}
    P\ts{A}^g&= r_{11}+r_{22} 
    =\dfrac12 
    \big[ 
        1+ (2\alpha^2 -1 ) f\ts{A}
    \big], \\
    P\ts{A}^e&= r_{33}+r_{44}
    =\dfrac12 
    \big[ 
        1- (2\alpha^2 -1 ) f\ts{A}
    \big], \\
    P\ts{B}^g&= r_{11}+r_{33} \notag \\
    &=\dfrac{1}{2}
    \big[ 
        1 + (2\alpha^2 -1) f\ts{B} \cos (2\Theta)
    \big] \notag \\
    &\hspace{1cm}
    -\alpha \sqrt{1-\alpha^2} f\ts{B} \sin (2\Theta) \sin \vartheta, \\
    P\ts{B}^e&=
    r_{22}+r_{44} \notag \\
    &=\dfrac{1}{2}
    \big[ 
        1 - (2\alpha^2 -1) f\ts{B} \cos (2\Theta)
    \big] \notag \\
    &\hspace{1cm}
    +\alpha \sqrt{1-\alpha^2} f\ts{B} \sin (2\Theta) \sin \vartheta . 
\end{align}
Notice that the probabilities in $\rho\ts{A}$ only depend on the initial state $\alpha$ and $f\ts{A}$, while for $\rho\ts{B}$, probabilities depend also on the commutator $\Theta$ and the phase $\vartheta$. 
This can be understood by recalling our assumption that Alice always turns on the detector first, i.e., $t(\tau\ts{A,0}) \leq t(\tau\ts{B,0})$. 
Thus, Alice cannot be influenced by Bob but Bob could be affected by Alice by receiving signals. 
In fact, if there is no signal at all ($\Theta=0$) then Bob's probabilities take the same form as Alice's. 

The entropy $S(\rho\ts{AB})$ is computed by 
\begin{align}
    S(\rho\ts{AB})&=-\sum_{i=1}^4 p_i \ln p_i,
\end{align}
where $p_i$ are the eigenvalues of $\rho\ts{AB}$, 
\begin{align}
    p_1&=\dfrac{ r_{11} + r_{44} }{2}
    + \sqrt{ \kako{ \dfrac{ r_{11} - r_{44} }{2} }^2 + |r_{14}|^2 }, \\
    p_2&=\dfrac{ r_{11} + r_{44} }{2}
    - \sqrt{ \kako{ \dfrac{ r_{11} - r_{44} }{2} }^2 + |r_{14}|^2 }, \\
    p_3&=\dfrac{ r_{22} + r_{33} }{2}
    + \sqrt{ \kako{ \dfrac{ r_{22} - r_{33} }{2} }^2 + |r_{23}|^2 }, \\
    p_4&=\dfrac{ r_{22} + r_{33} }{2}
    - \sqrt{ \kako{ \dfrac{ r_{22} - r_{33} }{2} }^2 + |r_{23}|^2 }. 
\end{align}
$S(\rho\ts{AB})$ is also used to measure entanglement between the field and the detectors.

\section{Results}
\label{sec: results}

\begin{figure}[tp]
    \centering
    \includegraphics[width=\columnwidth]{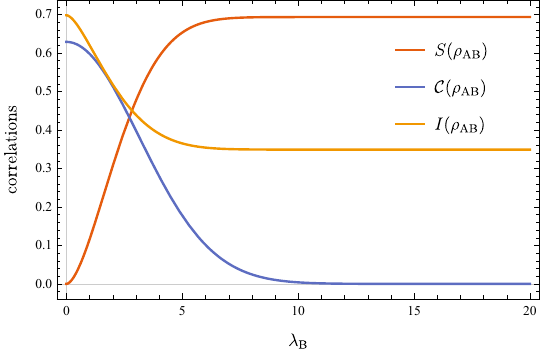}\\
    \caption{
    The trade-off property among the correlations $S(\rho\ts{AB}), \mathcal{C}(\rho\ts{AB})$, and $I(\rho\ts{AB})$. 
    Here $\alpha=1/3, \eta/\sigma=1$. 
    $\lambda\ts{B}=0$ represents the initial correlations, and $\lambda\ts{B}\gg 1$ corresponds to the extremely noisy scenario.}
    \label{fig:Single tradeoff}
\end{figure}

\begin{figure*}[tp]
    \centering
    \includegraphics[width=\textwidth]{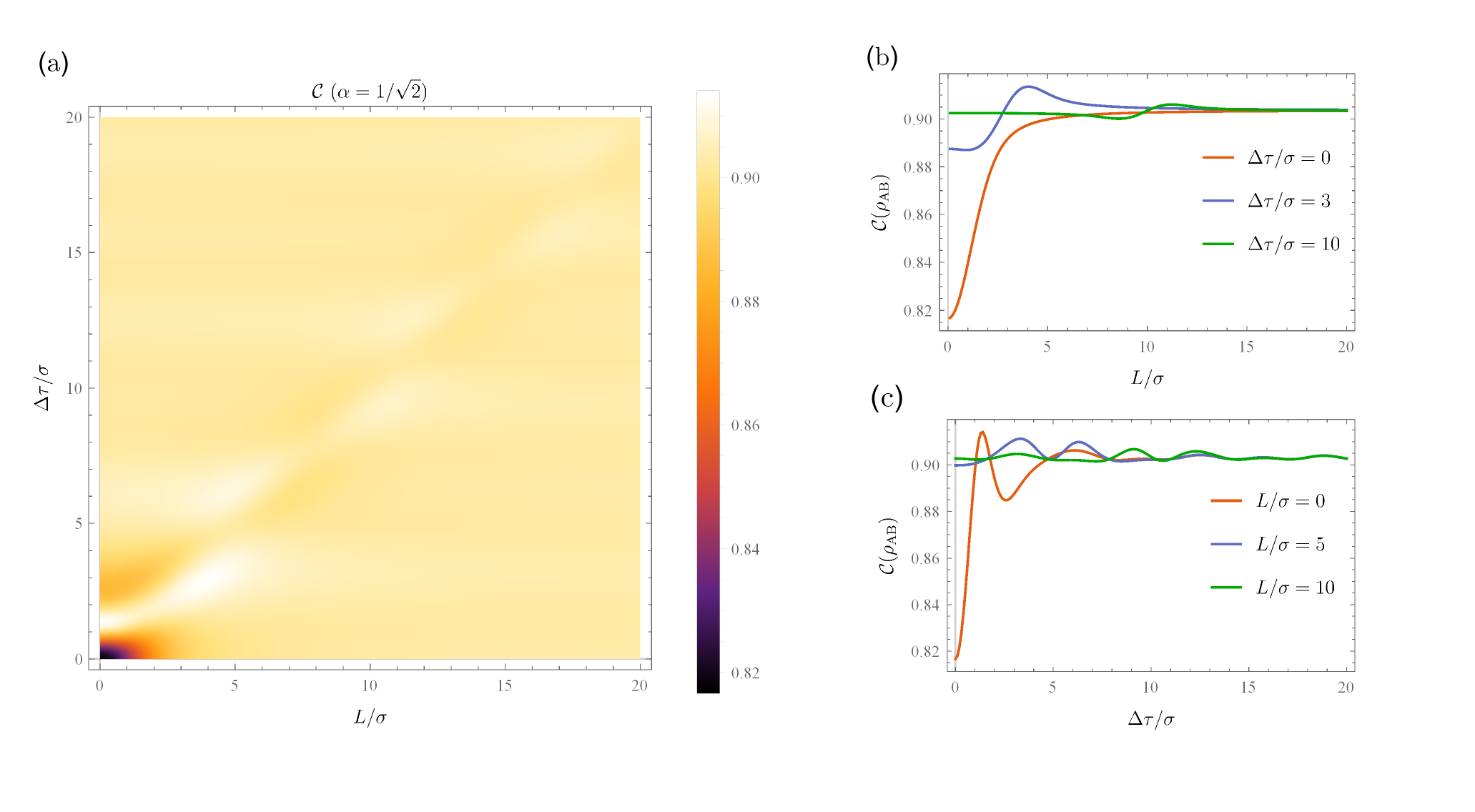}\\
    \caption{
    (a) Concurrence as a distribution of the detector separation $L/\sigma$ and switching difference $\Delta \tau/\sigma$. 
    Here we choose $\alpha=1/\sqrt{2}, \theta=0, \Omega \sigma=1, \lambda=1, \eta/\sigma=1$. 
    (b) Concurrence as a function of $L/\sigma$ at fixed $\Delta \tau/\sigma=0, 3$, and 10. 
    The value of $\mathcal{C}$ near $L=\Delta \tau$ is disturbed due to signalling from Alice to Bob. 
    (c) Concurrence as a function of $\Delta \tau/\sigma$ at fixed $L/\sigma=0, 5$, and 10. }
    \label{fig:Conc density max}
\end{figure*}

\subsection{Single detector interacting with the field}
\label{subsec:Single detector interacts with the field}
To begin with, let us consider the scenario where we prepare the initial state \eqref{eq:state 0011} and then only Bob interacts with the field. 
Since there no signalling effect comes into  play, we can examine how the initial entanglement is affected purely by the field. 

We set $\lambda\ts{A}=0$, which leads to $f\ts{A}=1$ and $\Theta=\omega=0$; thereby the elements \eqref{eq:r11 full}-\eqref{eq:r23 full} reduce to
\begin{subequations}
\begin{align}
    &r_{11}=
    \dfrac{\alpha^2}{2} ( 1 + f\ts{B} ),  \\
    &r_{22}
    =\dfrac{\alpha^2}{2} ( 1 - f\ts{B} ) , \\
    &r_{33}=
    \dfrac{1-\alpha^2}{2} ( 1 - f\ts{B} ) , \\
    &r_{44}=
    \dfrac{1-\alpha^2}{2} ( 1 + f\ts{B} ) , \\
    &r_{14}
    =
    \dfrac12 \alpha \sqrt{1-\alpha^2} (1+ f\ts{B} ) e^{-\ii \theta}, \\
    &r_{23} 
    =\dfrac{1}{2} \alpha \sqrt{1-\alpha^2}
    (1- f\ts{B}  ) e^{-\ii \theta} . 
\end{align}
\end{subequations}
Note that all the elements are independent of position in a spacetime. 

Concurrence $\mathcal{C}(\rho\ts{AB})$ and the terms in mutual information $I(\rho\ts{AB})$ are then
\begin{align}
    \mathcal{C}(\rho\ts{AB})&= 2 \alpha \sqrt{1-\alpha^2} f\ts{B}, \\
    S(\rho\ts{A})&= -\alpha^2 \ln \alpha^2 - (1-\alpha^2) \ln (1-\alpha^2), \\
    S(\rho\ts{B})&=- 
    \dfrac{ 1 + (2\alpha^2 -1) f\ts{B} }{2}
    \ln 
    \dfrac{ 1 + (2\alpha^2 -1) f\ts{B} }{2} \notag \\
    &\hspace{5mm}- 
    \dfrac{ 1 - (2\alpha^2 -1) f\ts{B} }{2}
    \ln 
    \dfrac{ 1 - (2\alpha^2 -1) f\ts{B} }{2}, \\
    S(\rho\ts{AB})&=
    - \dfrac{1+f\ts{B}}{2} \ln \dfrac{1+f\ts{B}}{2} 
    - \dfrac{1-f\ts{B}}{2} \ln \dfrac{1-f\ts{B}}{2}.
\label{SAB}    
\end{align}
Observe that concurrence is proportional to $f\ts{B} \in (0, 1]$, and it is non-increasing, namely, the detectors cannot gain entanglement after Bob interacts with the field, regardless of the value of $\alpha$. 
In fact, the stronger the field coupling, the greater the degradation of the initial
entanglement, which vanishes in the strong coupling regime as $\lambda\ts{B} \to \infty$. 
Furthermore, if the detectors are initially separable ($\alpha=0,1$) then
$\mathcal{C}(\rho\ts{AB})=0$ for all $\lambda\ts{B}$. 

Regarding the von Neumann entropies, $S(\rho\ts{A})$ is constant because Alice never interacts with the field, whereas Bob's detector will  obviously be affected.  
We see from \eqref{SAB} that the entanglement $S(\rho\ts{AB})$ between the field and the detectors only depends on $f\ts{B}$ and  not on $\alpha$. 
Regardless of the initial state, the entanglement between the field and the joint system of two detectors is the same.

\begin{figure*}[tp]
    \centering
    \includegraphics[width=\textwidth]{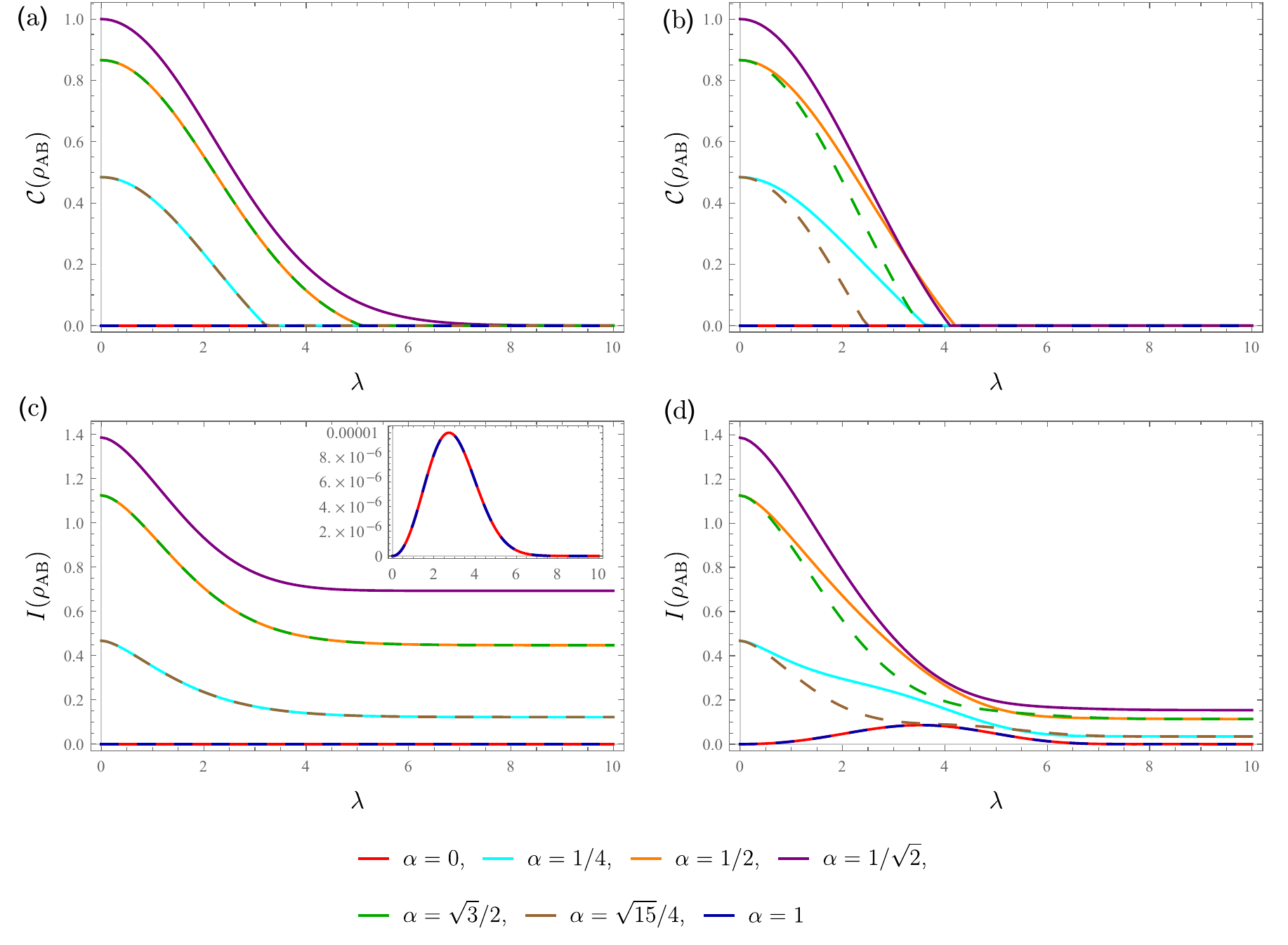}\\
    \caption{
    Concurrence (a), (b), and mutual information (c), (d) as a function of coupling strength $\lambda$ when the detectors are (a), (c) spacelike $L/\sigma=10, \tau\ts{A,0}=\tau\ts{B,0}=0$ and (b), (d) lightlike $L/\sigma=\Delta \tau/\sigma=1$, with $\theta=0, \Omega\sigma=1, \eta/\sigma=1$ for all cases. 
    Even though entanglement cannot be harvested, mutual information can be. 
    Also, lightlike separated detectors have different degradation properties with different $\alpha$ although the initial amount of entanglement is the same.}
    \label{fig:Conc Mutual coupling}
\end{figure*}

Notice that concurrence and mutual information have a trade-off property with the field-detectors entanglement $S(\rho\ts{AB})$ as shown in Fig. \ref{fig:Single tradeoff}. 
We find $\mathcal{C}(\rho\ts{AB})$ and $I(\rho\ts{AB})$ are maximum when $S(\rho\ts{AB})$ is minimum, and vice versa. 
This phenomenon can be thought of as the leakage of initial detector correlations to the quantum field. 

Although the concurrence vanishes at large $\lambda\ts{B}$,  Fig. \ref{fig:Single tradeoff} indicates that mutual information  is always nonzero. 
By setting $\lambda\ts{B} \to \infty$, $f\ts{B}$ becomes 0, and we get $S(\rho\ts{B})=S(\rho\ts{AB})=\ln 2$. 
Hence the mutual information in the strong coupling regime becomes $I(\rho\ts{AB})= S(\rho\ts{A})$, which is nonzero unless $\alpha=0,1$. 
This is exactly half the value of the initial mutual information, which can be evaluated by substituting $\lambda\ts{B}=0$, i.e., $f\ts{B}=1$, yielding $I(\rho\ts{AB})=2 S(\rho\ts{A})$. 
This is similar to situations in which  initially entangled modes become unentangled from the viewpoint of a uniformly accelerating observer in Minkowski spacetime  in the large acceleration limit \cite{PhysRevLett.95.120404},  or to a static observer in a spherically symmetric black hole spacetime with infinite Hawking temperature, while mutual information is degraded to half of its initial value \cite{PhysRevA.77.024302, PhysRevD.78.065015, WANG2009186}.

\subsection{Both detectors interacting with the field}
\label{subsec:Two detectors interact with the field}

We now consider two detectors interacting with the field. 
In what follows, all the values are in units of the typical Gaussian smearing width $\sigma$.

\begin{figure*}[tp]
    \centering
    \includegraphics[width=\textwidth]{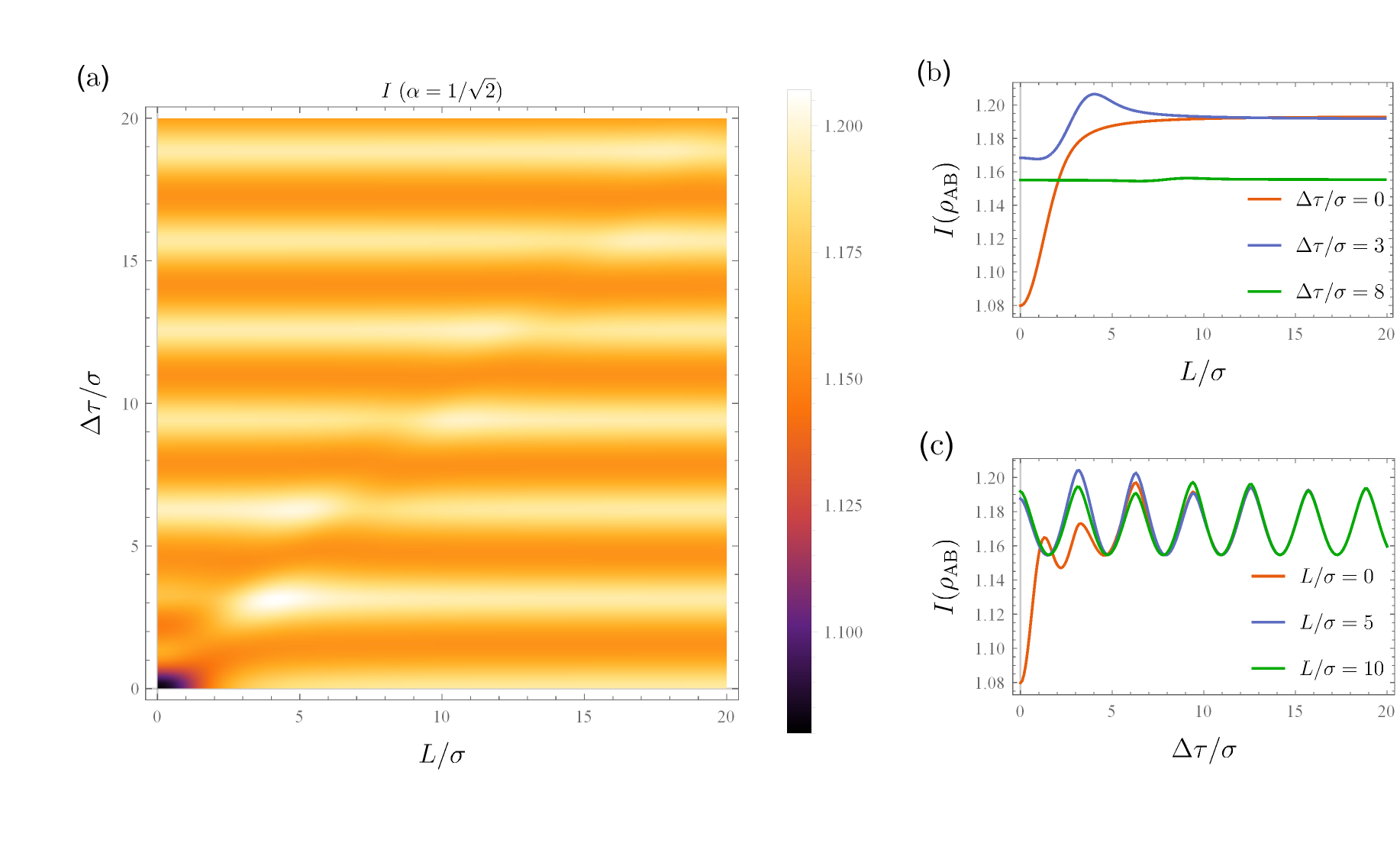}\\
    \caption{
    (a) Mutual information as a distribution of the detector separation $L/\sigma $ and switching difference $\Delta \tau/\sigma $. 
    Here we choose $\alpha=1/\sqrt{2}, \theta=0, \Omega \sigma=1, \lambda=1$, and $\eta/\sigma=1$. 
    (b) Mutual information as a function of $L/\sigma$ at fixed $\Delta \tau/\sigma=0, 3$, and 8. 
    (c) Mutual information as a function of $\Delta \tau/\sigma $ at fixed $L/\sigma=0, 5$, and 10.}
    \label{fig:Mutual density max}
\end{figure*}

\subsubsection{Entanglement degradation}

To begin with, let us look at the concurrence in a spacetime diagram. 
Figure \ref{fig:Conc density max} shows the concurrence when maximally entangled detectors are initially prepared. 
Figure \ref{fig:Conc density max}(a) is a density plot of $\mathcal{C}(\rho\ts{AB})$ in terms of the detector separation $L/\sigma$ and the switching difference $\Delta \tau/\sigma$, with $\theta=0, \Omega \sigma=1, \lambda=1, \eta/\sigma=1$. 
Given that the concurrence of the maximally entangled state is $\mathcal{C}(\rho\ts{AB})=1$, one immediately notices that the entanglement is degraded under all circumstances. 
Entanglement experiences its greatest degradation  for small values of $L/\sigma$ and $\Delta\tau/\sigma$, but asymptotes to about 90\% of its original value for large values of these quantities. 
We also notice that there is a pattern in the figure; the amount of entanglement changes periodically with $\Delta \tau/\sigma$ due to the phase factor $\vartheta$. 
This is more evident in Fig. \ref{fig:Conc density max}(c), which is a plot of concurrence as a function of $\Delta \tau/\sigma$ with fixed detector separation $L/\sigma$. 

Another characteristic can be seen when detectors are nearly lightlike separated, with  $L=\Delta \tau$. 
In this case the concurrence shows a distorting oscillation due to the exchange of quanta. 
This is clear from  Fig. \ref{fig:Conc density max}(b), where concurrence is plotted as a function of separation $L/\sigma$ with fixed $\Delta \tau/\sigma$. 
The concurrence is distorted in a region near $L=\Delta \tau$, but   becomes constant for large separation $L/\sigma$. 
This suggests that communication affects the amount of entanglement degradation. 

Let us now examine how field interactions affect the initial entanglement. 
Figures \ref{fig:Conc Mutual coupling}(a) and (b) depict concurrence as a function of the coupling strength $\lambda$ when detectors are (a) spacelike separated $L/\sigma=10, \tau\ts{A,0}/\sigma=\tau\ts{B,0}/\sigma=0$ and (b) lightlike separated $L/\sigma=\Delta \tau/\sigma=1$. 
In both cases, we set $\theta=0, \Omega \sigma=1$, and $\eta/\sigma=1$. 

Both scenarios show that entanglement is non-increasing,  vanishing at some value of $\lambda$. 
Since $\lambda=0$ implies that there is no interaction between the field and the detectors, the value of $\mathcal{C}(\rho\ts{AB})$ at $\lambda=0$ is the initial amount of entanglement. 
Note that initially separable detectors ($\alpha=0,1$) cannot harvest entanglement from the vacuum even if they are communication-assisted. 
This is consistent with earlier studies  \cite{Simidzija.Nonperturbative,Simidzija2018no-go} that found   delta-switched detectors cannot harvest entanglement from the field in a coherent state as well as a vacuum state.

We observe that communication modifies  degradation; even if the  detectors have the same amount of initial entanglement,   degradation   changes under swapping $\alpha$ and $\sqrt{1-\alpha^2}$ in the lightlike separated case. 
Recall that our initial entangled state was given in \eqref{eq:state 0011}. 
The initial concurrence  $\mathcal{C}(\rho\ts{AB})=2 \alpha \sqrt{1-\alpha^2}$ does not change under $\alpha \to \sqrt{1-\alpha^2}$. 
This is also true for spacelike separated detectors [Fig. \ref{fig:Conc Mutual coupling}(a)] but not for lightlike separated ones [Fig. \ref{fig:Conc Mutual coupling}(b)]. 

This effect can be verified by looking at the elements of the density matrix \eqref{eq:r11 full}-\eqref{eq:r23 full}. 
Since $2\alpha^2 -1$ changes its sign under $\alpha \to  \sqrt{1-\alpha^2}$, all elements will be modified. 
However, if the detectors cannot communicate, i.e., $\Theta=0$, then $r_{14}$ and a product $r_{22} r_{33}$ are invariant under such a transformation. 
Therefore, concurrence $\mathcal{C}(\rho\ts{AB})$ changes under $\alpha \to \sqrt{1-\alpha^2}$ when $\Theta \neq 0$. 
This kind of phenomenon has also been seen when one detector is noninertial \cite{PhysRevA.77.024302, PhysRevD.78.065015, WANG2009186}.

\subsubsection{Mutual information}

We will now look at mutual information $I(\rho\ts{AB})$. 
We first show in Fig. \ref{fig:Mutual density max} how $I(\rho\ts{AB})$ behaves in a spacetime when the detectors are initially maximally entangled. 
The behavior is similar to that of concurrence; it has periodic phase dependence and there is distortion of $I(\rho\ts{AB})$ when the detectors can communicate. 
In addition, the coupling dependence shown in Figs. \ref{fig:Conc Mutual coupling}(c) and (d) are also similar to the concurrence case in Figs. \ref{fig:Conc Mutual coupling}(a) and (b); signalling modifies the degradation property under $\alpha \to  \sqrt{1-\alpha^2}$. 

There are two notable departures from the behavior of the concurrence. 
One is the ability to harvest mutual information. 
The other is that of nonvanishing correlation in the strong coupling limit $\lambda\to \infty$. 

\begin{figure}[tp]
    \centering
    \includegraphics[width=\columnwidth]{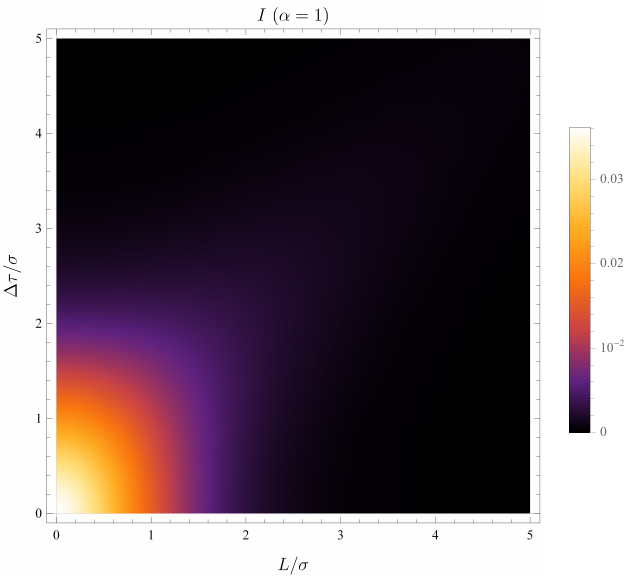}\\
    \caption{
    A density plot of mutual information when the detectors are initially a separable state. 
    Here we choose $\lambda=1, \Omega \sigma=1, \eta/\sigma=1$, and $\theta=0$. 
    Unlike entanglement, detectors can harvest mutual information from the vacuum using delta switching.}
    \label{fig:Mutual density}
\end{figure}

As shown in Figs. \ref{fig:Conc Mutual coupling}(c) and (d) with $\alpha=0$ and 1, detectors can harvest mutual information from the vacuum, even when the detectors cannot communicate with each other. 
Figure \ref{fig:Mutual density} depicts the mutual information in terms of detector separation when they are initially in their  ground states $\ket{g\ts{A} g\ts{B}}$. 
One can numerically confirm that setting $\alpha=0,1$ and $\Theta=0$ still gives nonzero mutual information, indicating that noncommunicating detectors can harvest mutual information. 
Since entanglement is always zero, the content of this harvested correlation is classical correlation or nondistillable entanglement. 

\begin{figure}[tp]
    \centering
    \includegraphics[width=\columnwidth]{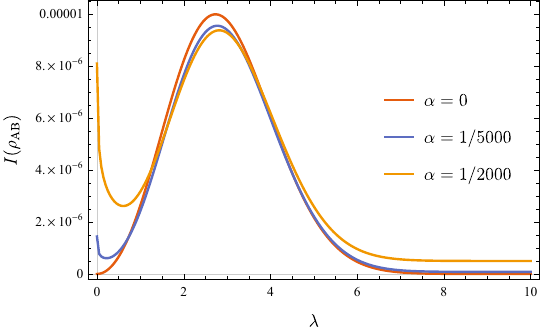}\\
    \caption{
    Mutual information acquisition can happen for initially separable, or weakly entangled detectors. 
    The red, blue and orange lines represent $\alpha=0, 1/5000$, and $1/2000$, respectively. 
    Here we set $\tau\ts{A,0}/\sigma=\tau\ts{B,0}/\sigma=0, L/\sigma=10, \Omega \sigma=1, \eta/\sigma=1$, and $\theta=0$. 
    For some range of $\lambda$, the detectors can gain mutual information from the vacuum compared to the initial state ($\lambda=0$). }
    \label{fig:Mutual acquisition}
\end{figure}

From the degradation point of view, the fact that detectors can harvest mutual information suggests that we can \textit{gain} some correlation by interacting with a field. 
In fact, slightly entangled detectors can gain mutual information from the vacuum, which is shown in Fig. \ref{fig:Mutual acquisition}. 
This figure shows that spacelike separated, slightly entangled detectors can acquire mutual information from the vacuum state; over some range of $\lambda$, the amount of mutual information is greater than the initial amount, which is given at $\lambda=0$. 
This suggests that there is a case where the anti-commutator $\omega$ enhances correlations between the detectors. 
Note that this kind of mutual information acquisition never happens if only one detector interacts with the field, as shown in Fig. \ref{fig:Single tradeoff}, because $\omega=0$. In addition,
  more strongly entangled detectors cannot gain correlation as shown in Fig. \ref{fig:Conc Mutual coupling}(c). 

The second characteristic we observe is that mutual information is nonvanishing at large $\lambda$ [Figs. \ref{fig:Conc Mutual coupling}(c), (d)] in contrast to entanglement [Figs. \ref{fig:Conc Mutual coupling}(a), (b)]. 
This  phenomenon has been previously observed in other contexts  \cite{PhysRevLett.95.120404, PhysRevA.74.032326, PhysRevA.77.024302, PhysRevD.78.065015, WANG2009186, PhysRevA.80.032315}. 
Since entanglement is zero as $\lambda \to \infty$ but nonzero for mutual information, the remaining correlations in the strong coupling regime are classical correlations and nondistillable entanglement.

\section{Conclusion}
\label{sec: conclusion}

We have considered entanglement degradation of two initially entangled, inertial UDW detectors coupled to a scalar field in a $(3+1)$-dimensional Minkowski spacetime, employing nonperturbative methods  \cite{Simidzija.Nonperturbative}. In contrast to situations in which  one of the UDW detectors is   accelerating,  we   only have considered inertial detectors. 
We are particularly interested in the effects the quantum vacuum has on detector entanglement (analogous to  the entanglement harvesting protocol), and how communication channels via the field affect this.

We investigated the amount of entanglement  for two cases:  a single detector interacting with the field after the entangled detectors are prepared, and both detectors interacts with the field. 
In each case the concurrence monotonically decreased and vanished for large coupling strength $\lambda$. 
In this sense, detectors with   delta switching experience entanglement degradation. 
Note that while signalling indeed affects the degradation property, it does not assist in acquiring entanglement.

Our results are complementary to those of Ref. \cite{Simidzija2018no-go}, in which a general no-go theorem was demonstrated that gave constraints on whether or not a separable bipartite system of finite dimension (e.g., detectors) could become entangled through local interactions with a source that was either finite or infinite dimensional (e.g., a quantum field). 
The constraints of this theorem do not apply in the present case, since our bipartite system of finite dimension (the two detectors) is initially entangled.

The overall behavior of mutual information is similar to concurrence, with the important exception that weakly entangled detectors can gain mutual information for some range of values of $\lambda$. As long as the detectors are initially entangled, the mutual information never vanishes at the strong coupling regime. 

Although the possibility of communication does not assist entanglement acquisition, it modifies the degradation properties of both entanglement and mutual information. 
We showed that two states $\alpha \ket{g\ts{A} g\ts{B}} + e^{\ii \theta} \sqrt{1-\alpha^2} \ket{e\ts{A}e\ts{B}}$ and $\sqrt{1-\alpha^2} \ket{g\ts{A} g\ts{B}} + e^{\ii \theta} \alpha \ket{e\ts{A}e\ts{B}}$, which are related by the transformation $\alpha \to  \sqrt{1-\alpha^2}$, experience different rates of degradation if and only if the two detectors can communicate, even though they possess the same amount of initial entanglement. 

We close by commenting on  what would happen if we employed the initial state \eqref{altpsi}.  
Such a state can be created from 
\eqref{eq:state 0011} by passing it through the bit-flip gate $\hat I\ts{A} \otimes \hat \sigma_x$, where $\hat \sigma_x$ is the Pauli-$x$ matrix acting on detector B. 
The resulting final density matrix of the detectors is 
\begin{align}
    \left[
    \begin{array}{cccc}
    r_{22}' &0 &0 &r_{23}'  \\
    0 &r_{11}' &r_{14}' &0  \\
    0 &(r_{14}')^* &r_{44}' &0  \\
    (r_{23}')^* &0 &0 & r_{33}'
    \end{array}
    \right],
\end{align}
where $r_{ij}'$ are equivalent to $r_{ij}$ in \eqref{eq:final density} with $\Omega\ts{B}\to -\Omega\ts{B}$. 
Therefore, the overall behavior of the entanglement and mutual information is essentially the same as $\alpha \ket{g\ts{A} g\ts{B}} + e^{\ii \theta} \sqrt{1-\alpha^2} \ket{e\ts{A}e\ts{B}}$ case.

\section*{Acknowledgment}

The authors thank Ahmed Shalabi and Laura J. Henderson for useful information regarding the nonperturbative calculations, and Erickson Tjoa, Eduardo Mart\'{i}n-Mart\'{i}nez and Petar Simidzija for discussion and clarification on the no-go theorem. 
K.G-Y. acknowledges the support from Keio University Global Fellowship. 
This work was supported in part by the Natural Sciences and Engineering Research Council of Canada and by Asian Office of Aerospace Research and Development Grant No. FA2386-19-1-4077.

\appendix
\section{THE ELEMENTS IN THE DENSITY MATRIX}
\label{appendix: density operator}
Let us consider two detectors in an $(n+1)$-dimensional Minkowski spacetime. 
The time-evolution operator $\hat U\ts{I}$ given in \eqref{eq: time-evolution operator} can be written as \cite{Simidzija.Nonperturbative}
\begin{align}
    \hat U\ts{I}&=
    \hat I\ts{A} \otimes \hat I\ts{B} \otimes \hat X_{(+,+)} 
    + \hat \mu\ts{A} \otimes \hat I\ts{B} \otimes \hat X_{(+,-)} \notag \\
    &+ \hat I\ts{A} \otimes \hat \mu\ts{B} \otimes \hat X_{(-,+)}
    + \hat \mu\ts{A} \otimes \hat \mu\ts{B} \otimes \hat X_{(-,-)},
\end{align}
where
\begin{align}
    \hat X_{(j,k)}\coloneqq 
    \dfrac{1}{4}
    \kako{
        e^{ \hat Y\ts{B} } + j e^{- \hat Y\ts{B} }
    }
    \kako{
        e^{ \hat Y\ts{A} } + k e^{- \hat Y\ts{A} }
    },
\end{align}
and we denote $\hat X_{(\pm , \pm)}=\hat X_{(\pm 1, \pm 1)}$. 

Assuming that the initial state is \eqref{eq:state 0011}, the final density matrix of the detectors takes the form
\begin{align}
    \rho\ts{AB}&=
        \alpha^2 \rho\ts{AB}^{gg} 
        +\alpha \sqrt{1-\alpha^2}e^{-\ii \theta}\rho\ts{AB}^{ge} \notag \\
        &+\alpha \sqrt{1-\alpha^2}e^{\ii \theta}\rho\ts{AB}^{eg}
        +(1-\alpha^2)\rho\ts{AB}^{ee},
\end{align}
where 
\begin{align}
    \rho\ts{AB}^{\alpha \beta} 
    &\coloneqq 
    \Tr_\phi
    \kagikako{
        \hat U\ts{I} 
        \kako{
            \ket{\alpha\ts{A}} \bra{\beta\ts{A}} \otimes
            \ket{\alpha\ts{B}} \bra{\beta\ts{B}} \otimes 
            \ket{0}\bra{0}
        }
        \hat U\ts{I}^\dag
    },\notag \\
    &\hspace{1cm}
    \alpha, \beta \in \{g,e\}.
\end{align}
For instance, $\rho\ts{AB}^{gg} $ reads
\begin{widetext}
\begin{align}
    \rho\ts{AB}^{gg}&=
    \Tr_\phi
    \kagikako{
        \hat U\ts{I} 
        \kako{
            \ket{g\ts{A}} \bra{g\ts{A}} \otimes
            \ket{g\ts{B}} \bra{g\ts{B}} \otimes 
            \ket{0}\bra{0}
        }
        \hat U\ts{I}^\dag
    } \\
    &= f_{(++++)} \ket{gg}\bra{gg}
    + f_{(+-++)}
    e^{ -\ii \Omega\ts{A} \tau\ts{A,0} } \ket{gg}\bra{eg} \notag \\
    &+ f_{(-+++)} e^{ -\ii \Omega\ts{B} \tau\ts{B,0} } \ket{gg}\bra{ge}
    + f_{(--++)}
    e^{ -\ii \Omega\ts{A} \tau\ts{A,0} } 
    e^{ -\ii \Omega\ts{B} \tau\ts{B,0} } \ket{gg}\bra{ee} \notag \\
    &+ f_{(+++-)}
    e^{ \ii \Omega\ts{A} \tau\ts{A,0} }
     \ket{eg}\bra{gg}
    +f_{(+-+-)}
     \ket{eg}\bra{eg} \notag \\
    &+ f_{(-++-)}
    e^{ \ii \Omega\ts{A} \tau\ts{A,0} } e^{ -\ii \Omega\ts{B} \tau\ts{B,0} }
    \ket{eg}\bra{ge} 
    + f_{(--+-)}
    e^{ -\ii \Omega\ts{B} \tau\ts{B,0} }
    \ket{eg}\bra{ee} \notag \\
    &+ f_{(++-+)}
    e^{ \ii \Omega\ts{B} \tau\ts{B,0} }
    \ket{ge}\bra{gg} 
    + f_{(+--+)}
    e^{ -\ii \Omega\ts{A} \tau\ts{A,0} } e^{ \ii \Omega\ts{B} \tau\ts{B,0} }
    \ket{ge}\bra{eg} \notag \\
    &+f_{(-+-+)}
    \ket{ge}\bra{ge} 
    +f_{(---+)}
    e^{ -\ii \Omega\ts{A} \tau\ts{A,0} }
    \ket{ge}\bra{ee} \notag \\
    &+f_{(++--)}
    e^{ \ii \Omega\ts{A} \tau\ts{A,0} } e^{ \ii \Omega\ts{B} \tau\ts{B,0} }
    \ket{ee}\bra{gg}
    +f_{(+---)}
    e^{ \ii \Omega\ts{B} \tau\ts{B,0} }
    \ket{ee}\bra{eg} \notag \\
    &+ f_{(-+--)}
    e^{ \ii \Omega\ts{A} \tau\ts{A,0} }
    \ket{ee}\bra{ge}
    + f_{(----)}
    \ket{ee}\bra{ee}, 
\end{align}
where 
\begin{align}
    f_{(jklm)}
    &\coloneqq 
    \bra{0} \hat X^\dag_{(j,k)} \hat X_{(l,m)} \ket{0} \\
    &=(1+jl+km+jklm)
    + k(1+jl) \bra{0} e^{ 2\hat Y\ts{A} } \ket{0}
    + m(1+jl) \bra{0} e^{ -2\hat Y\ts{A} } \ket{0} \notag \\
    & + l \bra{0} e^{ -\hat Y\ts{A} } e^{ -2\hat Y\ts{B} } e^{ \hat Y\ts{A} }
    + j \bra{0} e^{ -\hat Y\ts{A} } e^{ 2\hat Y\ts{B} } e^{ \hat Y\ts{A} }\ket{0}
    + kl \bra{0} e^{ \hat Y\ts{A} } e^{ -2\hat Y\ts{B} } e^{ \hat Y\ts{A} } \ket{0} \notag \\
    &+ jk \bra{0} e^{ \hat Y\ts{A} } e^{ 2\hat Y\ts{B} } e^{ \hat Y\ts{A} } \ket{0}
    + lm \bra{0} e^{ -\hat Y\ts{A} } e^{ -2\hat Y\ts{B} } e^{ -\hat Y\ts{A} }\ket{0}
    + jm \bra{0} e^{ -\hat Y\ts{A} } e^{ 2\hat Y\ts{B} } e^{ -\hat Y\ts{A} } \ket{0} \notag \\
    &+ klm \bra{0} e^{ \hat Y\ts{A} } e^{ -2\hat Y\ts{B} } e^{ -\hat Y\ts{A} } \ket{0}
    + kjm \bra{0} e^{ \hat Y\ts{A} } e^{ 2\hat Y\ts{B} } e^{ -\hat Y\ts{A} } \ket{0}.
\end{align}
$f_{(jklm)}$ can be simplified by using the Baker–Campbell–Hausdorff formula
\begin{align}
    e^{\hat A} e^{\hat B}=
    \exp 
    \kako{
        \hat A + \hat B
        +\dfrac{1}{2} [\hat A, \hat B]
        +\dfrac{1}{12} \big[ \hat A, [\hat A, \hat B] \big] 
        - \dfrac{1}{12} \big[ \hat B, [\hat A, \hat B] \big] 
        + \cdots
    }.
\end{align}
Given that $[\hat Y\ts{A}, \hat Y\ts{B}]=\ii \Theta$, we get 
\begin{align}
    e^{p\hat Y\ts{A}} e^{ q\hat Y\ts{B} } e^{ r\hat Y\ts{A} }
    &= e^{ \ii \Theta q(p-r)/2 } e^{ (p+r) \hat Y\ts{A} + q\hat Y\ts{B} },
    ~~~(p,q,r\in \mathbb{R})
\end{align}
Hence, the vacuum expectation values in $f_{(jklm)}$ will be reduced to one of the following \cite{Simidzija.Nonperturbative}.
\begin{align}
    &f_j\coloneqq
    \bra{0} e^{ 2\hat Y_j } \ket{0}
    =\exp 
    \kako{
        -\dfrac{1}{2} \int \dd^n k\,|\beta_j(\bm{k})|^2
    }, \\
    &f\ts{p}\coloneqq
     \bra{0} e^{ 2\hat Y\ts{B} } e^{ 2\hat Y\ts{A} } \ket{0}
    = \exp
    \kagikako{
        -\dfrac{1}{2} \int \dd^n k
        \kako{
            |\beta\ts{A}(\bm{k})|^2 + |\beta\ts{B}(\bm{k})|^2
            + 2\beta\ts{A}(\bm{k}) \beta\ts{B}^*(\bm{k})
        }
    }
    =f\ts{A} f\ts{B} e^{ \omega -2\ii \Theta }
    , \\
    &f\ts{m}\coloneqq
    \bra{0} e^{ 2\hat Y\ts{B} } e^{ -2\hat Y\ts{A} } \ket{0}
    =\exp
    \kagikako{
        -\dfrac{1}{2} \int \dd^n k
        \kako{
            |\beta\ts{A}(\bm{k})|^2 + |\beta\ts{B}(\bm{k})|^2
            - 2\beta\ts{A}(\bm{k}) \beta\ts{B}^*(\bm{k})
        }
    }
    =f\ts{A} f\ts{B} e^{ -\omega +2\ii \Theta },
\end{align}
where $\omega\coloneqq 2\bra{0} \{ \hat Y\ts{A}, \hat Y\ts{B} \}\ket{0}$ is the vacuum expectation value of the anti-commutator. 
Therefore, $f_{(jklm)}$ becomes 
\begin{align}
    f_{(jklm)}&=\dfrac{1}{16}
    \Bigkako{
        (1+jl+km+jklm)
        + (1+jl)(k+m) f\ts{A} \notag \\
        &\hspace{10mm}
        +\Bigkagikako{
            (l+jkm) e^{ 2\ii \Theta } + (j+klm) e^{ -2\ii \Theta }
        } f\ts{B}
        +\Bigkagikako{
            (jk+lm) e^\omega + (jm+kl) e^{-\omega}
        } f\ts{A}f\ts{B}
    }.
\end{align}
From this, notice that $f_{(+++-)}=f_{(++-+)}=f_{(+-++)}=f_{(-+++)}=f_{(---+)}=f_{(--+-)}=f_{(-+--)}=f_{(+---)}=0$. 
This fact simplifies $\rho\ts{AB}^{gg}$, and by using the basis $\{\, \ket{g\ts{A} g\ts{B}}, \ket{g\ts{A} e\ts{B}}, \ket{e\ts{A} g\ts{B}}, \ket{e\ts{A} e\ts{B}}\, \}$, we get 
\begin{align}
    \rho\ts{AB}^{gg}
    &=
    \left[
    \begin{array}{cccc}
    r_{11}^{gg} &0 &0 &r_{14}^{gg}  \\
    0 &r_{22}^{gg} &r_{23}^{gg} &0  \\
    0 &(r_{23}^{gg})^* &r_{33}^{gg} &0  \\
    (r_{14}^{gg})^* &0 &0 & r_{44}^{gg}
    \end{array}
    \right],
\end{align}
where 
\begin{align}
    r_{11}^{gg}&=f_{(++++)}, \\
    r_{14}^{gg}&=f_{(--++)} e^{ -\ii (\Omega\ts{A} \tau\ts{A,0} + \Omega\ts{B} \tau\ts{B,0}) }, \\
    r_{22}^{gg}&= f_{(-+-+)}, \\
    r_{23}^{gg}&= f_{(+--+)} e^{ -\ii (\Omega\ts{A} \tau\ts{A,0} - \Omega\ts{B} \tau\ts{B,0}) }, \\
    r_{33}^{gg}&= f_{(+-+-)}, \\
    r_{44}^{gg}&= f_{(----)}. 
\end{align}
With $\rho\ts{AB}^{ge}, \rho\ts{AB}^{eg},\rho\ts{AB}^{ee}$, the total density matrix becomes 
\begin{align}
    \rho\ts{AB}
    &= 
    \left[
    \begin{array}{cccc}
    r_{11} &0 &0 &r_{14}  \\
    0 &r_{22} &r_{23} &0  \\
    0 &r_{23}^* &r_{33} &0  \\
    r_{14}^* &0 &0 & r_{44}
    \end{array}
    \right], \label{eq:max entangled density matrix}
\end{align}
where 
\begin{align}
    &r_{11}=\alpha^2 f_{(++++)} 
    +\alpha \sqrt{1-\alpha^2}e^{-\ii\theta} f_{(--++)} 
    e^{ \ii(\Omega\ts{A} \tau\ts{A,0} + \Omega\ts{B} \tau\ts{B,0}) } \notag \\
    &\hspace{10mm}
    + \alpha \sqrt{1-\alpha^2}e^{\ii\theta} 
    f_{(++--)} e^{ -\ii(\Omega\ts{A} \tau\ts{A,0} + \Omega\ts{B} \tau\ts{B,0}) }
    + (1-\alpha^2) f_{(----)}, \\
    &r_{14}=\alpha^2 f_{(--++)} e^{ -\ii (\Omega\ts{A} \tau\ts{A,0} + \Omega\ts{B} \tau\ts{B,0}) }
    +\alpha \sqrt{1-\alpha^2}e^{-\ii\theta} f_{(++++)} \notag \\
    &\hspace{10mm}
    + \alpha \sqrt{1-\alpha^2}e^{\ii\theta} f_{(----)} e^{ -\ii 2(\Omega\ts{A} \tau\ts{A,0} + \Omega\ts{B} \tau\ts{B,0}) } 
    + (1-\alpha^2) f_{(++--)}e^{ -\ii (\Omega\ts{A} \tau\ts{A,0} + \Omega\ts{B} \tau\ts{B,0}) }, \\
    &r_{22}=\alpha^2 f_{(-+-+)}
    +\alpha \sqrt{1-\alpha^2}e^{-\ii\theta} f_{(+--+)} e^{ \ii(\Omega\ts{A} \tau\ts{A,0} + \Omega\ts{B} \tau\ts{B,0}) } \notag \\
    &\hspace{10mm}
    + \alpha \sqrt{1-\alpha^2}e^{\ii\theta} f_{(-++-)}e^{ -\ii (\Omega\ts{A} \tau\ts{A,0} + \Omega\ts{B} \tau\ts{B,0}) }
    + (1-\alpha^2) f_{(+-+-)}, \\
    &r_{23}=\alpha^2 f_{(+--+)} e^{ -\ii (\Omega\ts{A} \tau\ts{A,0} - \Omega\ts{B} \tau\ts{B,0} ) }
    +\alpha \sqrt{1-\alpha^2}e^{-\ii\theta} f_{(-+-+)} e^{ \ii 2\Omega\ts{B}\tau\ts{B,0} } \notag \\
    &\hspace{10mm}
    + \alpha \sqrt{1-\alpha^2}e^{\ii\theta} f_{(+-+-)}e^{ -\ii 2\Omega\ts{A}\tau\ts{A,0} }
    + (1-\alpha^2) f_{(-++-)} e^{ -\ii (\Omega\ts{A} \tau\ts{A,0} - \Omega\ts{B} \tau\ts{B,0}) }, \\
    &r_{33}=\alpha^2 f_{(+-+-)}
    +\alpha \sqrt{1-\alpha^2}e^{-\ii\theta} f_{(-++-)} e^{ \ii(\Omega\ts{A} \tau\ts{A,0} + \Omega\ts{B} \tau\ts{B,0}) } \notag \\
    &\hspace{10mm}
    + \alpha \sqrt{1-\alpha^2}e^{\ii\theta} f_{(+--+)}e^{ -\ii(\Omega\ts{A} \tau\ts{A,0} + \Omega\ts{B} \tau\ts{B,0}) }
    + (1-\alpha^2) f_{(-+-+)}, \\
    &r_{44}=\alpha^2 f_{(----)}
    +\alpha \sqrt{1-\alpha^2}e^{-\ii\theta} f_{(++--)} 
    e^{ \ii(\Omega\ts{A} \tau\ts{A,0} + \Omega\ts{B} \tau\ts{B,0}) } \notag \\
    &\hspace{10mm}
    + \alpha \sqrt{1-\alpha^2}e^{\ii\theta} f_{(--++)} e^{ -\ii(\Omega\ts{A} \tau\ts{A,0} + \Omega\ts{B} \tau\ts{B,0}) }
    + (1-\alpha^2) f_{(++++)}, \\
    &f_{(\pm \pm \pm \pm )}=
    \dfrac{1}{4}
    \Bigkagikako{
        1 \pm f\ts{A} \pm f\ts{B}\cos (2\Theta) + f\ts{A}f\ts{B} \cosh \omega
    },\\
    &f_{(\pm \pm \mp \mp)}=
    \mp \dfrac14 f\ts{B}
    \Bigkagikako{
        \ii \sin (2\Theta) \mp f\ts{A} \sinh \omega
    }, \\
    &f_{(\pm \mp \pm \mp )}=
    \dfrac14 
    \Bigkagikako{
        1\mp f\ts{A} \pm f\ts{B}\cos (2\Theta) - f\ts{A}f\ts{B}\cosh \omega
    }, \\
    &f_{(\pm  \mp  \mp  \pm )}=
    \mp \dfrac14 f\ts{B}
    \Bigkagikako{
        \ii \sin (2\Theta) \pm f\ts{A} \sinh \omega
    }. 
\end{align}
These expressions still can be simplified. 
For simplicity, let us define $\vartheta=\Omega\ts{A} \tau\ts{A,0} + \Omega\ts{B} \tau\ts{B,0} -\theta$. 
Let us consider $r_{11}$ as an example. 
By denoting $f_{(\pm \pm \pm \pm )}= \frac14 (P \pm Q)$ where $P\coloneqq 1+ f\ts{A} f\ts{B} \cosh \omega$ and $Q\coloneqq f\ts{A} + f\ts{B} \cos (2\Theta)$, we get the following. 
\begin{align}
    r_{11}&= \alpha^2 f_{(++++)} 
    + \alpha \sqrt{1-\alpha^2} f_{(--++)} e^{ \ii \vartheta } 
    + \alpha \sqrt{1-\alpha^2} f_{(++--)} e^{-\ii \vartheta}
    +(1-\alpha^2) f_{(----)} \notag \\
    &= \dfrac{1}{4}\alpha^2 (P + Q) + \dfrac{1}{4}(1-\alpha^2) (P-Q) 
    +2\alpha \sqrt{1-\alpha^2} \text{Re} [ f_{(--++)} e^{\ii \vartheta} ] \\
    &=\dfrac{1}{4} 
    \big[ P + (2\alpha^2 -1)Q \big]
    +2 \alpha \sqrt{1-\alpha^2}
    \text{Re}
    \kagikako{
        \dfrac{f\ts{B}}{4}[ \ii \sin(2\Theta) + f\ts{A}\sinh \omega ] e^{\ii \vartheta}
    } \\
    &=\dfrac{1}{4} 
    \big[ P + (2\alpha^2 -1)Q \big]
    +\dfrac{f\ts{B}}{2} \alpha \sqrt{1-\alpha^2}
    \big[ 
        f\ts{A} \sinh \omega \cos \vartheta - \sin (2\Theta) \sin \vartheta
    \big]. 
\end{align}
Via a similar procedure, we obtain the final form \eqref{eq:r11 full}-\eqref{eq:r23 full}. 

Finally, let us derive $f_j, \Theta$ and $\omega$ given in \eqref{eq:eff}, \eqref{eq:Theta2} and \eqref{eq:omega in 4D flat}. 
We choose the Gaussian smearing function \eqref{eq:Gaussian smearing} so that its Fourier transformation is 
\begin{align}
    \tilde{F}(\bm{k})= \dfrac{1}{\sqrt{ (2\pi)^n }} e^{ -|\bm{k}|^2 \sigma^2/4 }.
\end{align}
Assuming $n=3$, $f_j$ can be evaluated by substituting $\tilde{F}(\bm{k})$ to $\beta_j(\bm{k})$ \eqref{eq:beta}.
\begin{align}
    f_j&=
    \exp
    \kako{
        -\dfrac{1}{2} \int \dd^n k\, |\beta_j(\bm{k})|^2
    } \notag \\
    &=
    \exp
    \kako{
        -2 \lambda_j^2 \eta_j^2 
        \int \dfrac{ \dd^n k }{ (2\pi)^n 2 |\bm{k}| }\,e^{ -|\bm{k}|^2 \sigma^2/2 }
    } \\
    &=\exp
    \kako{
        -2 \lambda_j^2 \eta_j^2 
        \dfrac{ 4\pi }{(2\pi)^3 2}
        \int_0^\infty \dd |\bm{k}|\,|\bm{k}| e^{ -|\bm{k}|^2 \sigma^2/2 }
    } \\
    &=\exp
    \kako{
        -
        \dfrac{ \lambda_j^2 \eta_j^2 }{ 2\pi^2 \sigma^2 }
    },
\end{align}
where we have used $\dd^3k=|\bm{k}|^2 \sin \theta \dd |\bm{k}| \dd \theta \dd \phi$ in the spherical coordinates. 

Now let us assume that two detectors are identical, namely, $\lambda \coloneqq \lambda\ts{A}=\lambda\ts{B}, \eta \coloneqq \eta\ts{A}=\eta\ts{B}$, and $\Omega \coloneqq \Omega\ts{A}=\Omega\ts{B}$. 
$\Theta$ can be computed as 
\begin{align}
    \Theta &=
    \dfrac{\ii}{4} 
    \int \dd^3 k
    \Bigkagikako{
        \beta\ts{A}^*(\bm{k}) \beta\ts{B}(\bm{k})
        -
        \beta\ts{A}(\bm{k}) \beta\ts{B}^*(\bm{k})
    } \notag \\
    &= \dfrac{\ii}{4} \dfrac{4\lambda^2 \eta^2}{2 (2\pi)^3} 
    \int \dd^3k \dfrac{ e^{ -\kk^2 \sigma^2/2 } }{\kk} 
    \kagikako{
        e^{ \ii \kk (\tau\ts{B,0}-\tau\ts{A,0}) }
        e^{ -\ii \bm{k}\cdot (\bm{x}\ts{B}-\bm{x}\ts{A}) }
        -\text{c.c.}
    }\\
    &=\dfrac{\ii \lambda^2 \eta^2}{2(2\pi)^3}
    \int_0^\infty \dd \kk \, \kk^2 \dfrac{ e^{ -\kk^2 \sigma^2/2 } }{\kk} 
    \int_0^\pi \dd \theta\,\sin \theta 
    \kagikako{
        e^{ \ii \kk \Delta \tau } e^{ -\ii \kk L \cos \theta } - \text{c.c.}
    }
    \int_0^{2\pi} \dd \phi \\
    &=\dfrac{\ii \lambda^2 \eta^2 (2\pi)}{2(2\pi)^3}
    \int_0^\infty \dd \kk \, \kk e^{ -\kk^2 \sigma^2/2 }
    \dfrac{ 2 \sin (\kk L) }{ \kk L }
    \kako{
        e^{ \ii \kk \Delta \tau } - e^{ -\ii \kk \Delta \tau }
    }\\
    &=\dfrac{- 2\lambda^2 \eta^2 }{(2\pi)^2 L}
    \int_0^\infty \dd \kk \, e^{ -\kk^2 \sigma^2/2 }
    \sin (\kk L)
    \sin (\kk \Delta \tau)\\
    &=\dfrac{\lambda^2 \eta^2}{ 4\pi^2 L \sigma } 
    \sqrt{ \dfrac{\pi}{2} } 
    \kako{
        e^{ -(\Delta \tau + L)^2/2\sigma^2 } - e^{ -(\Delta \tau - L)^2/2\sigma^2 } 
    }.
\end{align}
Also for $\omega$,
\begin{align}
    \omega&=
    -\dfrac{1}{2} \int \dd^3 k
    \Bigkagikako{
        \beta\ts{A}^*(\bm{k}) \beta\ts{B}(\bm{k})
        +
        \beta\ts{A}(\bm{k}) \beta\ts{B}^*(\bm{k})
    } \notag \\
    &=-\dfrac{1}{2} \dfrac{ 4 \lambda^2 \eta^2 }{ 2 (2\pi)^3 }
    \int \dd^3 k\, \dfrac{ e^{ -|\bm{k}|^2 \sigma^2/2 } }{ |\bm{k}| } 
    \Bigkako{
        e^{ \ii|\bm{k}| \Delta \tau } e^{ -\ii \bm{k}\cdot \Delta \bm{x} } + \text{c.c.}
    } \\
    &=-\dfrac{1}{2} \dfrac{ 4 \lambda^2 \eta^2 }{ 2 (2\pi)^3 }
    \int_0^\infty \dd |\bm{k}|\,|\bm{k}|^2 \dfrac{ e^{ -|\bm{k}|^2 \sigma^2/2 } }{ |\bm{k}| } 
    \int_0^\pi \dd \theta\,\sin \theta 
    \Bigkako{
        e^{ \ii|\bm{k}| \Delta \tau } e^{ -\ii |\bm{k}| L \cos \theta } + \text{c.c.}
    } 
    \int_0^{2\pi} \dd \phi \\
    &=-\dfrac{1}{2} \dfrac{ 4 \lambda^2 \eta^2 (2\pi) }{ 2 (2\pi)^3 }
    \int_0^\infty \dd |\bm{k}|\,|\bm{k}| e^{ -|\bm{k}|^2 \sigma^2/2 }
    \dfrac{2 \sin (\kk L) }{\kk L} 2\cos (\kk \Delta \tau)
    \\
    &=-\dfrac{\lambda^2 \eta^2}{ \sqrt{2} \pi^2 \sigma L }
    \kagikako{
        D^+ 
        \kako{
            \dfrac{ \Delta \tau + L }{ \sqrt{2}\sigma }
        }
        -
        D^+ 
        \kako{
            \dfrac{ \Delta \tau - L }{ \sqrt{2}\sigma }
        }
    },
\end{align}
where $D^+(x)$ is the Dawson function.

\end{widetext}

\bibliography{ref}

\end{document}